\documentclass[english]{appolb}
\usepackage{graphicx} 
\usepackage{arydshln} 
\usepackage{float} 

\usepackage{units} 
\usepackage{amsbsy}
\usepackage{babel}

\usepackage{amsmath}
\newcommand\Tstrut{\rule{0pt}{2.6ex}}       
\newcommand\Bstrut{\rule[-1.1ex]{0pt}{0pt}} 

\begin{document}
\title{Separation Between d$_{\unitfrac{5}{2}}$ and s$_{\unitfrac{1}{2}}$
Neutron Single Particle Strength in $^{15}$N%
\thanks{Presented at the XXXIII Mazurian Lakes Conference on Physics, Piaski,
Poland, September 1-7, 2013}$^,$\thanks{This work was supported by the U.S. %
Department of Energy and the U.S. National Science Foundation}}

\author{C. E. Mertin$^1$, D.D. Caussyn$^1$, A. M. Crisp$^1$, N. Keeley$^2$, %
 \\ K.W. Kemper$^1$, O. Momotyuk$^1$, B.T. Roeder$^1$, and A. Volya$^1$\\ %
\hfill \\ \begin{small}
$^1${\em Department of Physics, Florida State University, Tallahassee, %
Florida, 32306-4350, USA} \\ and \\ $^2${\em National Centre for %
Nuclear Research, ul.\ Andrzeja So\l tana 7, 05-400 Otwock, Poland}
\end{small}}

\maketitle
\begin{abstract}
The separation between single particle levels in nuclei plays the dominant role 
in determining the location of the neutron drip line. The separation also
provides a test of current crossed shell model interactions if the 
experimental data is such that multiple shells are involved. The present work
uses the $^{14}$N(d, p)$^{15}$N reaction to extract the 2s$_{1/2}$, and
1d$_{5/2}$ total neutron single particle strengths and then compares these
results with a shell model calculation using a p-sd crossed shell interaction to
identify the J$^\pi$ of all levels in $^{15}$N up to 12.8 MeV in excitation.\\
\end{abstract}

\newpage

\section{Introduction}
The nucleus $^{15}$N was a source of early shell model studies \cite{Kur56}
because it has a large number of unnatural parity states that were seen as a
puzzle since it has two sd states around 5.3 MeV and then a
multiplet of sd states around 7.7 MeV. In addition, around 9 MeV in excitation,
negative parity states appear mixed with additional positive parity states. A 
detailed measurement of the $^{14}$N(d, p) reaction at bombarding energies of 7, 
8, and 9 MeV when coupled to an earlier work that focused on the ground state
strengths in the p shell provided single particle strengths for all states in
$^{15}$N up to 10 MeV in excitation that displayed enough experimental yield to
extract angular distributions \cite{Sch67,Phi69}. These experimental results
were then used by Lie and Engeland \cite{Lie76} to test their weak coupling
shell model calculations. Their shell model calculations were compared to the
experimentally observed levels in $^{15}$N \cite{Ajz70} with the aim of
assigning J$^{\pi}$ values to the levels whose presence were known but with
ambiguous spin parity values. Also, there was the possibility that these
calculations could lead to the finding of new levels in this nucleus. 

\begin{figure}[H]
\centerline{
\includegraphics[width=12.5cm]{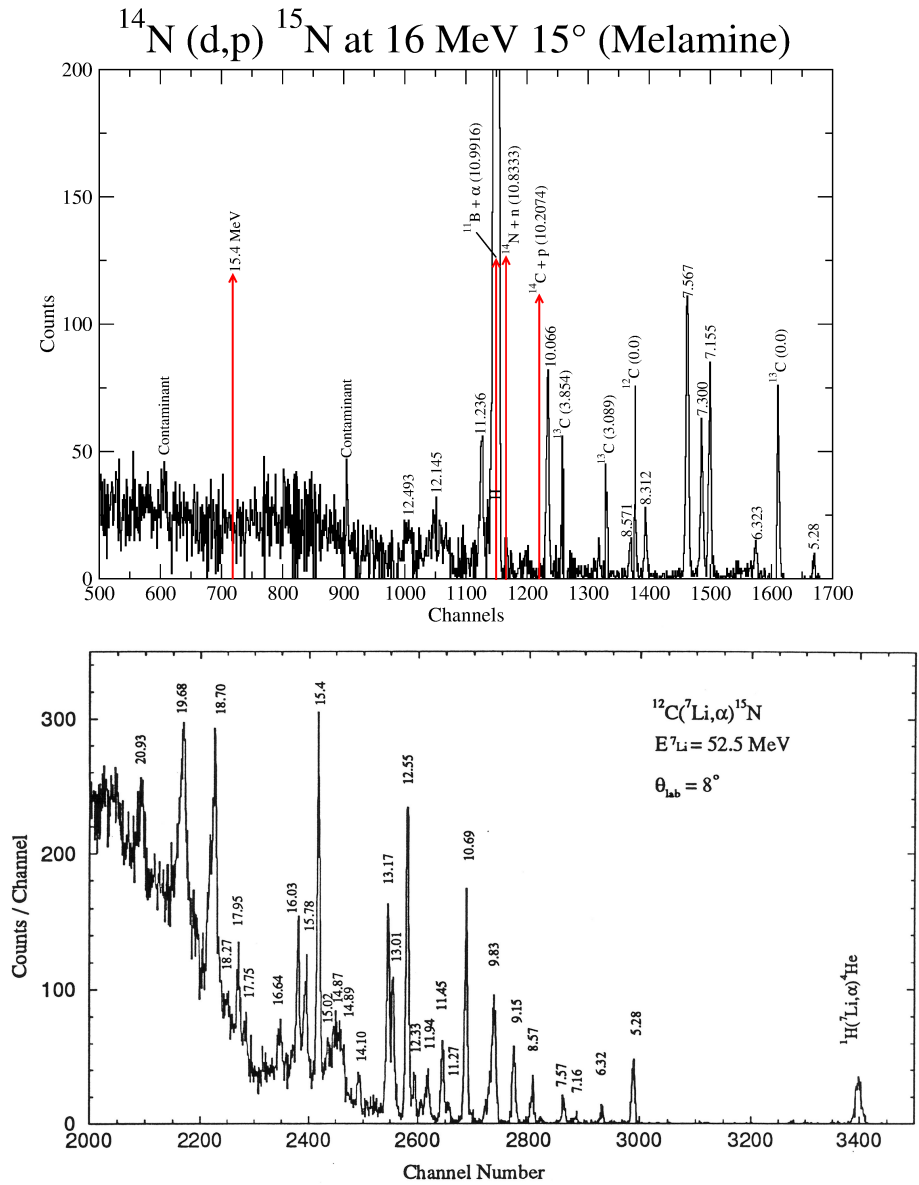}}
\caption{These spectra show the differences between the single particle transfer
reaction $^{14}$N(d, p) vs that of the three particle transfer reaction 
$^{12}$C($^7$Li, $\alpha$) \cite{Lee99}. In the three particle transfer, the
higher energy states are more strongly populated than those at lower energy
states. The inverse is true for the single particle transfer, with the lower
energy states being much stronger than the higher energy states.}
\label{Fig:3Trans}
\end{figure}

At about the same time as the calculations of Lie and Engeland were taking
place, the use of three particle transfer reactions to locate mirror states in
light nuclei was realized \cite{Bin71,Hol74} including a study to determine the
location of mirror states in $^{15}$N$ - $$^{15}$O \cite{Any74}. These and other
studies \cite{Any74} discovered a rich spectrum with quite different population
strengths from those observed in the (d, p) reaction. While the (d, p) reaction
seemed to have little or no strength above 11 MeV, multi-particle transfer
reactions populated states up to 20 MeV or more in $^{15}$N. The present work
uses the higher energy deuteron bombarding energy of 16 MeV to explore this
higher energy excitation region with the $^{14}$N(d, p) reaction. Figure
\ref{Fig:3Trans} is a comparison of spectra from the present $^{14}$N(d,
p)$^{15}$N work and that from previously measured spectra for $^{12}$C($^7$Li,
$\alpha$)$^{15}$N \cite{Lee99}. Note the presence of strong isolated peaks in
($^{7}$Li, $\alpha$) around 12, 13, and 15 MeV in excitation that are absent
in the single neutron transfer spectrum.

In addition to new experimental data, shell model calculations were carried out
that used an unrestricted p-sd valence space and an interaction Hamiltonian
taken from the work of Utsuno and Chiba \cite{Uts01}. In addition to giving a
high quality description of the $0 \hbar \omega$ and $1 \hbar \omega$ states,
this interaction was found to describe well multiparticle correlations in this
mass range as well as alpha clustering features \cite{VolIASEN}. Time dependent
and traditional shell model techniques were used for the calculations with the
code COSMO \cite{Vol09}.

\newpage

\section{Experiment}
	The measurement of the $^{14}$N(d, p) reaction used the FSU
tandem-linac accelerator system to produce the 16 MeV deuteron beam that
bombarded a melamine (C$_3$H$_6$N$_6$) target. A silicon surface barrier $\Delta
E-E$ telescope was used to collect the d elastic and p transfer data at 
laboratory angles of 10-35$^\circ$ in steps of 5$^\circ$. A fixed monitor
detector was used to determine the relative angular distribution for the
elastic and proton transfer data and to provide a continuous monitor of the
target thickness since melamine has a low melting point and it is possible that
target material could boil away during the experiment. The beam current was kept
below 15pnA and no target changes were observed during the experiment. The
elastic scattering data in the movable detector were matched to previously
measured elastic deuteron-carbon and nitrogen \cite{Jar74} scattering to get the
absolute cross section values. Based on the absolute cross section of the
published d-$^{12}$C,$^{14}$N scattering data and the statistics gathered in the 
present run, the absolute cross section is accurate to $\pm 20\%$. Since the 
emphasis in the current work was on the investigation of the single neutron 
strength of high lying states in $^{15}$N, the detector gains were set so that 
no data were taken for the ground state transition. Previously published data 
exist \cite{Sch67} and these were used in the current analysis. As can be seen 
in the typical spectrum shown in Figure \ref{Fig:3Trans}, the carbon and hydrogen 
in the target provide significant overlaps with the states of interest. To make 
certain that no peak coming from a contaminant was misidentified as one in 
$^{15}$N, a scattering chamber was filled with pure nitrogen gas and spectra 
were obtained at the same angles as with the melamine target. While these 
spectra have worse energy resolution due to the beam and ejectile straggling 
and energy loss of the particle in the gas, they showed that no states populated 
in $^{15}$N were missed or covered up by the presence of the hydrogen and carbon 
in the melamine target.

\section{Results}
	The measured angular distributions of the $^{15}$N states were  
compared to calculated Finite-range Distorted Wave Born Approximation (DWBA) 
angular distributions, performed using the code {\sc Fresco} \cite{Tho88}, 
to extract the relative single particle spectroscopic strengths. Deuteron and 
proton optical model potentials were taken from Phillips and Jacobs \cite{Phi69}. 
The deuteron bound state wave function was calculated using the Reid soft core 
potential \cite{Rei68} and the transferred neutron was bound to the $^{14}$N 
core in a well of Woods-Saxon shape of radius $1.25 \times {\mathrm A}^{1/3}$ fm 
and diffuseness 0.65 fm with a spin-orbit component of the usual Thomas form and
depth of 6 MeV and the same radius and diffuseness parameters as the central
part; the depth of the central part was adjusted to give the correct binding
energy. The calculations were performed using the post form of the DWBA and
included the full complex remnant term.

\begin{figure}[h!b]
\centerline{
\includegraphics[width=6.5cm]{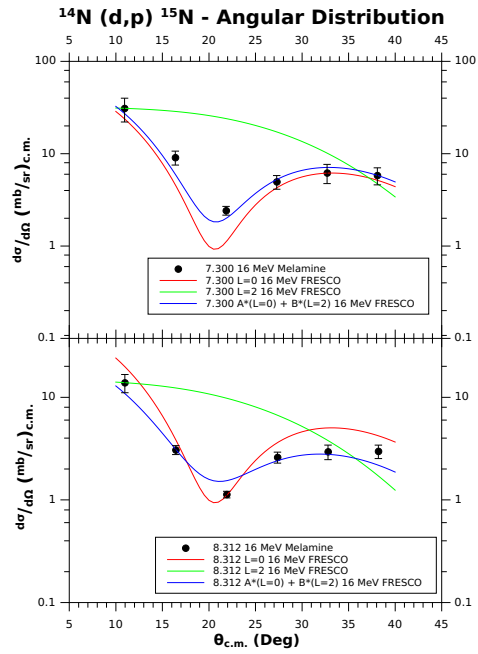}}
\caption{Angular distributions of L=0 transfer dominant states identified in 
the $^{14}$N(d, p) reaction. }
\label{Fig:L0}
\end{figure}

\begin{figure}[h!t]
\centerline{
\includegraphics[width=6.5cm]{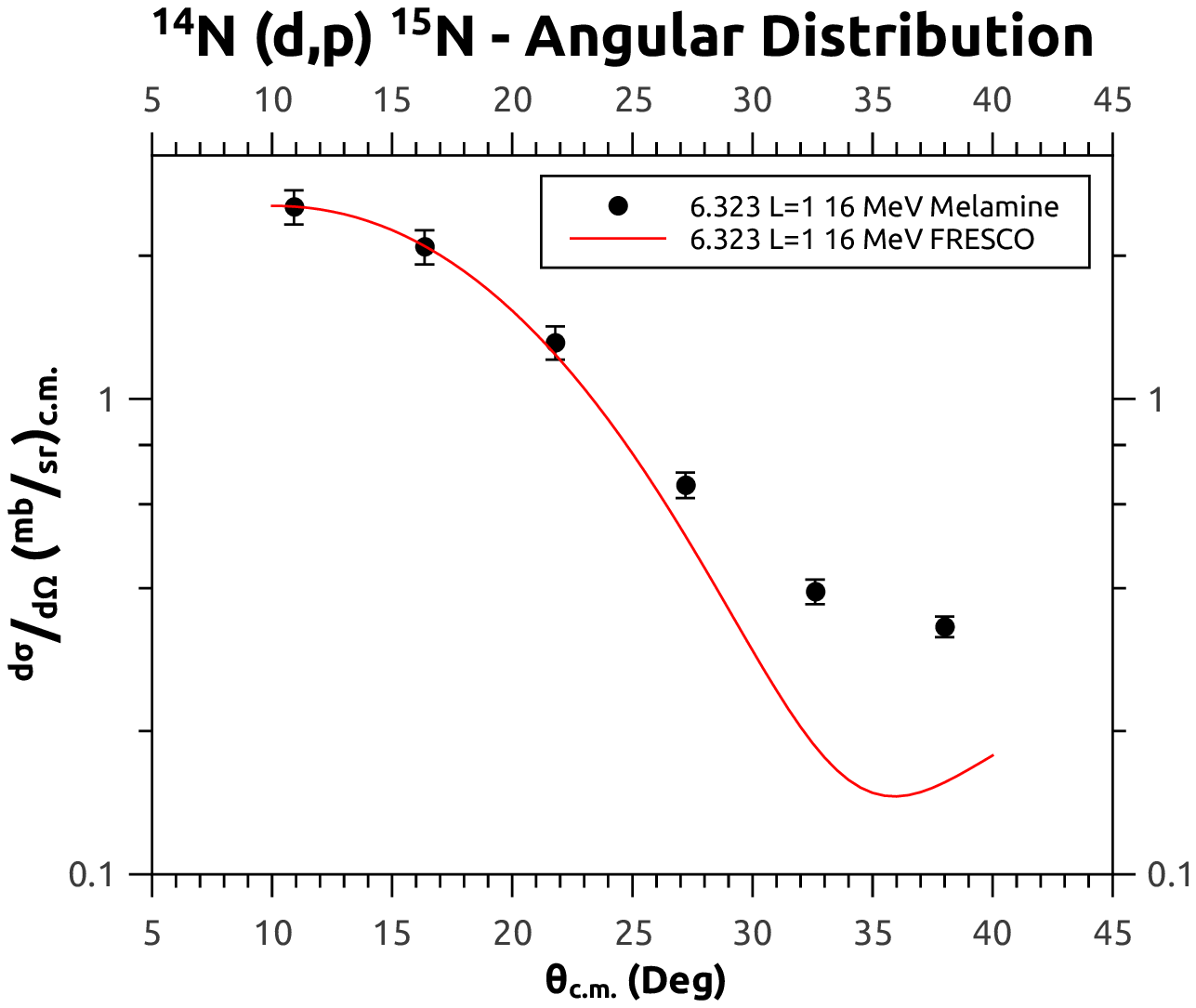}}
\caption{Angular distribution of L=1 dominant state identified in the
$^{14}$N(d, p) reaction. }
\label{Fig:L1}
\end{figure}

\begin{figure}[H]
\centerline{
\includegraphics[width=12.5cm]{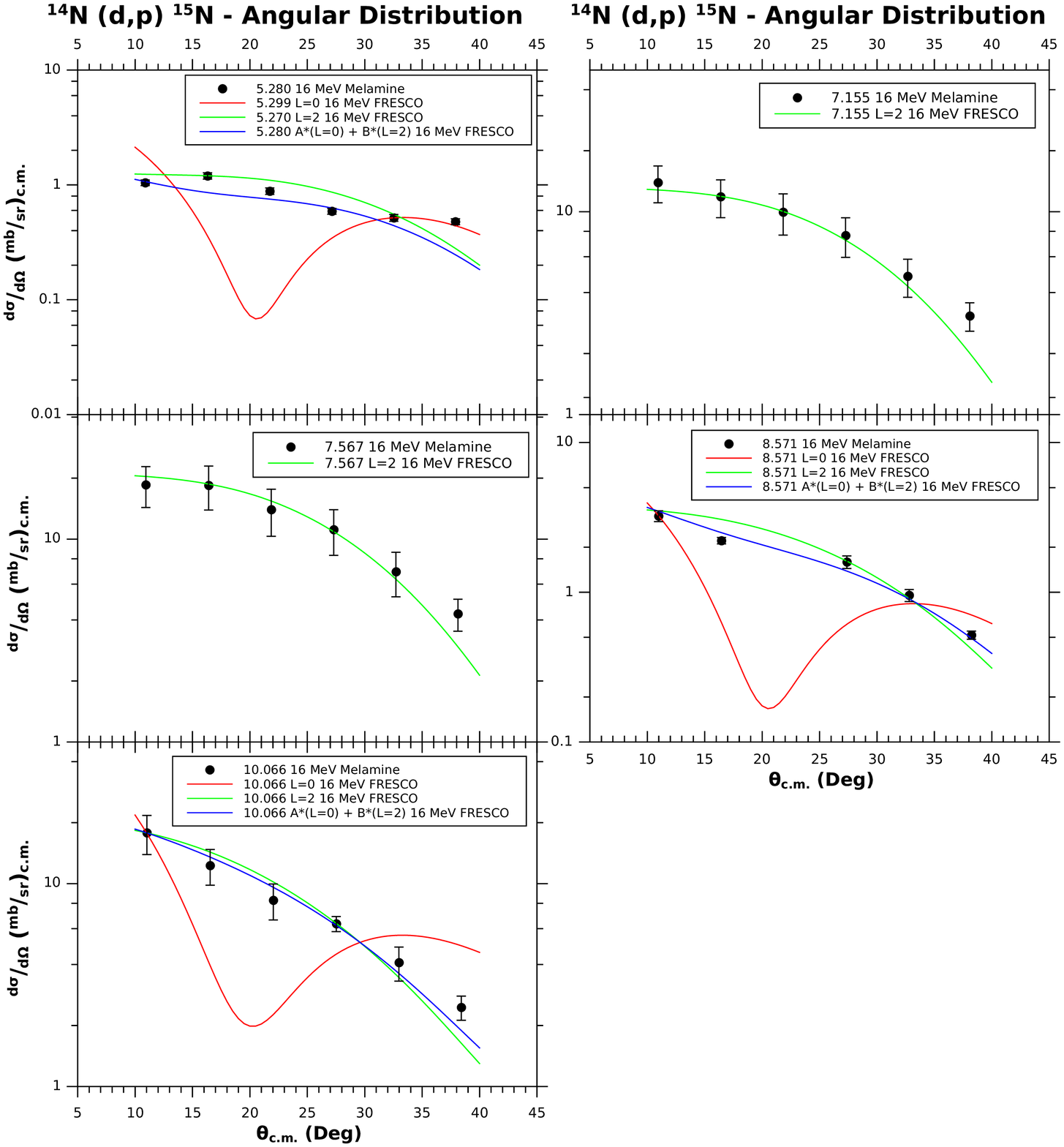}}
\caption{Angular distributions of L=2 dominant states identified in the
$^{14}$N(d, p) reaction. }
\label{Fig:L2}
\end{figure}

 Figure \ref{Fig:L0} shows the angular distributions that we have identified as
being described as primarily arising from an L=0 transfer, and Figure
\ref{Fig:L2} are those arising from  a dominant L=2 transfer. Also, shown in
Figure \ref{Fig:L1} is the L=1 transfer to the 6.323 MeV state. The
spectroscopic factors for the transfers were determined by normalizing the DWBA
calculations to the experimental data. Table \ref{table:values} compares the
neutron spectroscopic factors extracted from the current work with those from
the previous work of Phillips and Jacobs \cite{Phi69} as well as the results of
the shell model calculations.

\begin{table}[H]
\caption{Spectroscopic Factors for the $^{14}$N(d, p)$^{15}$N Reaction}
\centerline{
\begin{tabular}{c c c c c c c}
\hline\hline 
\textbf{$\boldsymbol{^{15}}$N State} & \textbf{J$\boldsymbol{^\pi}$} &
\textbf{Nucleus State} & \textbf{L Value} & \textbf{16 MeV} &
\textbf{Theoretical} & \textbf{9 MeV}\footnotemark[1]\Tstrut\Bstrut \\ \hline 
\textbf{5.280}\footnotemark[2] & $\unitfrac{1}{2}^+$  & s$_{\unitfrac{1}{2}}$ &
L=0 & 0.02 $\pm$ 0.004 & 0.39 & $<$ 0.05\Tstrut \\
& $\unitfrac{5}{2}^+$  & d$_{\unitfrac{5}{2}}$  & L=2 & 0.10 $\pm$ 0.020 & 0.11
& $<$ 0.05 \\ 
\textbf{6.323} & $\unitfrac{3}{2}^{-}$  & p$_{\unitfrac{3}{2}}$ & L=1 & 0.20
$\pm$ 0.040 & 0.02 & 0.10 $\pm$ 0.02 \\
\textbf{7.155} & $\unitfrac{5}{2}^{+}$ & d$_{\unitfrac{5}{2}}$ & L=2 & 0.98
$\pm$ 0.020 & 0.65 & 0.88 $\pm$ 0.03 \\
\textbf{7.300} & $\unitfrac{3}{2}^+$  & s$_{\unitfrac{1}{2}}$ & L=0 & 1.10 $\pm$
0.024 & 0.72 & 0.89 $\pm$ 0.04 \\
&  & d$_{\unitfrac{5}{2}}$ & L=2 & 0.11 $\pm$ 0.008 & 0.07 & 0.07 $\pm$ 0.05 \\
\textbf{7.567} & $\unitfrac{7}{2}^{+}$ & d$_{\unitfrac{5}{2}}$ & L=2 & 1.05
$\pm$ 0.020 & 0.73 & 0.87 $\pm$ 0.01 \\
\textbf{8.312} & $\unitfrac{1}{2}^{+}$  & s$_{\unitfrac{1}{2}}$ & L=0 & 0.89
$\pm$ 0.004 & 0.65 & 1.02 $\pm$ 0.04 \\
&  & d$_{\unitfrac{3}{2}}$ & L=2 & 0.27 $\pm$ 0.030 & 1.74E-004 & $<$ 0.09 \\
\textbf{8.571} & $\unitfrac{3}{2}^{+}$  & s$_{\unitfrac{1}{2}}$ & L=0 & 0.04
$\pm$ 0.020 & 5.21E-006 & 0.02 $\pm$ 0.01 \\
&  & d$_{\unitfrac{5}{2}}$ & L=2 & 0.21 $\pm$ 0.026 & 0.13 & 0.12 $\pm$ 0.03 \\
\textbf{10.066} & $\unitfrac{3}{2}^{+}$ & s$_{\unitfrac{1}{2}}$ & L=0 & 0.15
$\pm$ 0.013 & 0.04 & 0.32 $\pm$ 0.08 \\
&  & d$_{\unitfrac{5}{2}}$ & L=2 & 0.86 $\pm$ 0.008 & 0.55 & 0.48 $\pm$
0.08\Bstrut \\
\hline
\end{tabular}
}
\label{table:values}
\end{table}

\footnotetext[1]{Data extracted from \cite{Phi69}}
\footnotetext[2]{The energy resolution was not sufficient to separate the 5.270
and 5.299 MeV states of $^{15}$N.}

As can be seen there is good agreement between the values extracted by Phillips
and Jacobs \cite{Phi69} in their 9 MeV study and the present work so that the
normal concern of the possible dependence on bombarding energy due to non-direct
neutron transfer is not a factor in describing the strong transitions. However, a
major difference observed in the $^{14}$N(d, p) spectrum in Figure
\ref{Fig:3Trans} is the population of states around 9 MeV in excitation, which are 
much more strongly populated at the low bombarding energy of 3 MeV \cite{Amo72} 
than in the present work showing that for these states, compound processes are 
very important at the lower energy so that these low energy data cannot be used 
to extract meaningful spectroscopic factors.

The total 2s$_{\unitfrac{1}{2}}$ and 1d$_{\unitfrac{5}{2}}$ neutron strength is
spread over numerous states because the ground state of $^{14}$N has a spin
and parity of 1$^+$. This means that the neutron strength of the 2s$_{\unitfrac{1}{2}}$
level will be found in final states of spin parity $\unitfrac{1}{2}^+$ and
$\unitfrac{3}{2}^+$ while that of the 1d$_{\unitfrac{5}{2}}$ level  will be
found in $\unitfrac{3}{2}^+$, $\unitfrac{5}{2}^+$, and $\unitfrac{7}{2}^+$. For the 
$\unitfrac{3}{2}^+$ states which can be populated by both L=0 and L=2 transfer, 
a best fit to the angular distribution was formed. The summed spectroscopic strengths 
were found by forming the sums of the respective single particle strengths. 
Then the energy weighted summed strengths, which can be found in Table \ref{table:centroids},
were calculated to determine the splitting between the s$_{{\unitfrac{1}{2}}}$ and 
d$_{{\unitfrac{5}{2}}}$ neutron orbits for the three sets of data.

\begin{table}[H]
\caption{Sum Rule and Centroid Locations for s$_{\unitfrac{1}{2}}$ and
d$_{\unitfrac{5}{2}}$ neutrons in $^{15}$N}
\centerline{
\begin{tabular}{c c c c}
\hline\hline 
& \textbf{16 MeV} & \textbf{Theoretical} & \textbf{9 MeV}\Tstrut\Bstrut \\
\hline 
\textbf{Sum Rule $\boldsymbol{\left(\text{s}_{\unitfrac{1}{2}}\ \otimes\
1^{+}\right)}$} &  &  & \Tstrut \\
\hfill$\boldsymbol{J^{\pi}=\unitfrac{1}{2}^{+}}$ & 0.910 & 0.760 & 1.230 \\
\hfill$\boldsymbol{J^{\pi}=\unitfrac{3}{2}^{+}}$ & 1.254 & 0.656 & 1.040 \\
\textbf{Sum Rule $\boldsymbol{\left(\text{d}_{\unitfrac{5}{2}}\ \otimes\
1^{+}\right)}$} &  &  & \\
\hfill$\boldsymbol{J^{\pi}=\unitfrac{3}{2}^{+}}$ & 1.174 & 0.823 & 0.600 \\
\hfill$\boldsymbol{J^{\pi}=\unitfrac{5}{2}^{+}}$ & 1.080 & 0.763 & 0.930 \\
\hfill$\boldsymbol{J^{\pi}=\unitfrac{7}{2}^{+}}$ & 1.050 & 0.731 & 0.870\Bstrut
\\
\hdashline
\textbf{Centroid $\boldsymbol{\left(\text{s}_{\unitfrac{1}{2}}\right)}$ } &
7.904 MeV & 7.835 MeV & 8.138 MeV\Tstrut \\
\textbf{Centroid $\boldsymbol{\left(\text{d}_{\unitfrac{5}{2}}\right)}$} & 8.068
MeV & 8.016 MeV & 7.918 MeV \\
$\boldsymbol{\Delta\left(\text{d}_{\unitfrac{5}{2}}-\text{s}_{\unitfrac{1}{2}}\right)}$ %
& 0.164 MeV & 0.181 MeV & -0.220 MeV\Bstrut \\
\hline
\end{tabular}
}
\label{table:centroids}
\end{table}

As Table \ref{table:centroids} shows, the location of the centroids for each of
the sets of data is roughly the same, and therefore independent of the bombarding
energy producing $^{15}$N. The summed strength over each neutron 
state is on the order of unity. This, in conjunction with the theoretical 
calculations, shows that all the major 2s$_{\unitfrac{1}{2}}$ and 1d$_{\unitfrac{5}{2}}$
single neutron particle strengths in $^{14}$N(d, p) were found.  

\section{Conclusion}
The present work shows that the extracted spectroscopic strengths at the higher
deuteron bombarding energy used here agree well with those previously extracted 
from the $^{14}$N(d, p) reaction at what was considered a low enough energy 
(9 MeV) that compound nucleus contributions might be important. However, the 
current agreement shows that the assumption by Phillips and Jacobs that the 
(d, p) reaction was direct at their lower energies is correct. The neutron 
energy centroids extracted here show that the neutron s$_{\unitfrac{1}{2}}$ and 
d$_{\unitfrac{5}{2}}$ orbits are roughly degenerate. This means that the 
energy splitting of these two orbits changes rapidly around $^{15}$N since in 
$^{13}$C the s$_{\unitfrac{1}{2}}$ orbit is 760 keV below the d$_{\unitfrac{5}{2}}$ orbit
 while in $^{17}$O, the s$_{\unitfrac{1}{2}}$ orbit is 870 keV above the 
d$_{\unitfrac{5}{2}}$ orbit. This work also demonstrates that modern p-sd shell 
model calculations reliably reproduce this trend as well as being able to 
predict total neutron orbit strengths in light nuclei. 


\newpage

\begin{thebibliography}{99}
\bibitem{Kur56}D. Kurath, {\it Phys.\ Rev.\/} {\bf 101}, 216 (1956); E. C.
Halbert and J.B. French, {\it Phys.\ Rev.\/} {\bf 105}, 1563 (1957); A. M. Lane,
{\it Rev.\ Mod.\ Phys.\/} {\bf 32}, 519 (1960).
\bibitem{Sch67}J.P. Schiffer {\it et al.}, {\it Phys.\ Rev.\/} {\bf 164}, 1274
(1967).
\bibitem{Phi69}G.W. Phillips and W.W. Jacobs, {\it Phys.\ Rev.\/} {\bf 184},
1052 (1969). 
\bibitem{Lie76}S. Lie and T. Engeland, {\it Nuc.\ Phys.\/} {\bf A267} 123 (1976)
and references in there.
\bibitem{Ajz70}F. Ajzenberg-Selove, {\it Nuc.\ Phys.\/} {\bf A152} 1 (1970).
\bibitem{Bin71}H. G. Bingham, H. T. Fortune, J. D. Garrett, and R. Middleton,
{\it Phys.\ Rev.\ Lett.\/} {\bf 26}, 1448 (1971).
\bibitem{Hol74}C. H. Holbrow, H. G. Bingham, R. Middleton, and J. D. Garrett,
{\it Phys.\ Rev.\/} {\bf 9}, 902 (1974).
\bibitem{Any74}N. Anyas-Weiss {\it et al.}, {\it Phys.\ Rep.\/} {\bf 12}, 201
(1974); H. G. Bingham, M. L. Halbert, D. C. Hensley, E. Newman, K. W. Kemper,
and L. A. Charlton, {\it Phys.\ Rev.\/} {\bf C11}, 1913 (1975).
\bibitem{Lee99}C. Lee, J. A. Liendo, P. D. Cathers, N. R. Fletcher, K. W.
Kemper, and P. L. Kerr, {\it Phys.\ Rev.\/} {\bf C60} 024317 (1999).
\bibitem{Uts01}Y. Utsuno and S. Chiba, {\it Phys.\ Rev.\/} {\bf C83} 021301(R)
(2001).
\bibitem{VolIASEN}A. Volya and Y.M. Tchuvil'sky, to be presented IASEN 2013,
http://iasen-2013.jinr.ru
\bibitem{Vol09}A. Volya, {\it Phys.\ Rev.\/} {\bf C79}, 044308 (2009);
http://cosmo.volya.net
\bibitem{Jar74}N. Jarmie, J. H. Jett, and R. J. Semper, {\it Phys.\ Rev.\/} {\bf
C10} 1748 (1974).
\bibitem{Tho88}I.J. Thompson, {\it Comput.\ Phys.\ Rep.\/} {\bf 7}, 167 (1988).
\bibitem{Amo72}A. Amokrane {\it et al.}, {\it Phys.\ Rev.\/} {\bf C6} 1934
(1988).
\bibitem{Rei68}R.V. Reid Jr., {\it Ann.\ Phys.\ (NY)} {\bf 50}, 441 (1968).
\end{thebibliography}
\end{document}